\documentclass[10pt,journal,twocolumn,nofonttune]{IEEEtran}
\usepackage{times}
\usepackage{amsmath}
\usepackage{bbm}
\usepackage{amssymb}
\usepackage{graphicx,dblfloatfix}
\usepackage{cite}
\usepackage{subeqnarray}
\usepackage{xcolor}
\usepackage{stmaryrd}
\usepackage{multirow}
\usepackage{algorithm}
\usepackage{algorithmic}
\usepackage{diagbox}
\usepackage{url}
\usepackage{pifont}
\usepackage{hyperref}
\usepackage{upgreek}
\usepackage{booktabs}

\newcommand{\nc}{\newcommand}
\nc{\bsm}{\boldsymbol}
\nc{\mbb}{\mathbb}
\nc{\mbs}{\mathbbmss}
\nc{\mbf}{\mathbf}
\nc{\mcl}{\mathcal}

\newcommand{\xiaowuhao}{\fontsize{9pt}{\baselineskip}\selectfont}

\begin{document}
\title{Cascaded Channel Estimation for Large Intelligent Metasurface Assisted Massive MIMO}
\author{Zhen-Qing He and Xiaojun Yuan\IEEEmembership{, Senior Member, IEEE} \vspace{-0.03cm}
\thanks{Z.-Q. He and X. Yuan are with the Center for Intelligent Networking and Communications, the National Key Laboratory of Science and
Technology on Communications, University of Electronic Science and Technology of China, Chengdu 611731, China (e-mail: zhenqinghe@uestc.edu.cn; xjyuan@uestc.edu.cn).}
}

\maketitle

\begin{abstract}
In this letter, we consider the problem of channel estimation for large intelligent metasurface (LIM) assisted massive multiple-input multiple-output (MIMO) systems. The main challenge of this problem is that the LIM integrated with a large number of low-cost metamaterial antennas can only passively reflect the incident signal by a certain phase shift, and does not have any signal processing capability. To deal with this, we introduce a general framework for the estimation of the transmitter-LIM and LIM-receiver cascaded channel, and propose a two-stage algorithm that includes a sparse matrix factorization stage and a matrix completion stage. Simulation results illustrate that the proposed method can achieve accurate channel estimation for LIM-assisted massive MIMO systems.
\end{abstract}
\begin{keywords}
Bilinear factorization, channel estimation, large intelligent metasurface, massive MIMO, matrix completion. \vspace{-3mm}
\end{keywords}

\section{Introduction}

\IEEEPARstart{M}{assive} multiple-input multiple-output (MIMO), as a promising technology for future wireless systems, has attracted growing research interest in both academia and industry over recent years. Although massive MIMO exhibits huge potentials to support a significantly large amount of mobile data traffic and wireless connections, implementing this system with large-scale antenna arrays in practice remains very challenging due to high hardware cost and increased power consumption. To achieve green and sustainable wireless networks, researchers have started looking into  energy efficient techniques to improve the system performance, ranging from the utilization of energy efficient hardware components to the design of green resource allocation and transceiver signal processing algorithms.

To reduce energy consumption and enhance communication quality in wireless networks, the large intelligent metasurface (LIM) \cite{liaskos2018new,hu2018beyond1, di2019smart}, a.k.a. the intelligent reflecting surface \cite{wu2018intelligent} or the reconfigurable intelligent surface \cite{huang2018large}, has been recently proposed as an innovative technology that conceptually goes beyond contemporary massive MIMO communications. Metamaterials, as an emerging technology known for its flexibility in manipulating electromagnetic waves, have found applications such as in radar \cite{sleasman2017experimental} and imaging \cite{wu2018range}, etc. As a potential application of the metamaterial in wireless communications, the LIM with integrated electronics retains almost all the advantages of massive MIMO such as allowing for an unprecedented focusing of energy that enables highly efficient wireless charging and remote sensing. Although traditional reflecting surfaces have a variety of applications in radar and satellite communications, their application in terrestrial wireless communication was not possible earlier. This is because these reflecting surfaces only had fixed phase shifters that could not adapt the induced phases with the time-varying channels which generally constitute the wireless propagation environments.

To achieve full potentials of the LIM-aided massive MIMO systems, the channel state information (CSI) between the base station (BS) and the LIM and between the LIM and the receiver is essential in reflect beamforming \cite{wu2018intelligent}, energy-efficient design \cite{huang2018large}, as well as in simultaneous passive beamforming and information transfer \cite{yan2019}. The main challenge of the CSI acquisition of the  LIM-assisted MIMO systems is that the LIM, unlike the BS or the receiver, only passively reflects the electromagnetic waves, and does not have any signal processing capability. By leveraging the programmable property of the LIM and the rank-deficient structure of the massive MIMO channel, we formulate the BS-LIM and LIM-receiver cascaded channel estimation as a combined sparse matrix factorization and matrix completion problem. To solve the problem, we present a two stage algorithm which includes the sparse matrix factorization stage and the matrix completion stage. The proposed two stage algorithm includes the bilinear generalized approximate message passing (BiG-AMP) \cite{parker2014bilinear} for sparse matrix factorization  and the Riemannian manifold gradient-based algorithm for matrix completion \cite{wei2016guarantees}. Numerical results demonstrate that our algorithm is able to achieve accurate channel estimation for LIM-assisted massive MIMO systems.

To the best of our knowledge, this is the first attempt to tackle the cascaded channel estimation problem for the LIM-assisted massive MIMO systems with all passive elements in the LIM. Notice that the authors in \cite{taha2019enabling} proposed a compressive sensing and training based deep learning approach for the LIM-assisted MIMO channel estimation. Nevertheless, in \cite{taha2019enabling} a few active antenna elements are implemented in the LIM to circumvent the challenging cascaded channel estimation problem by using conventional channel estimation techniques. In addition, we note that the BiG-AMP algorithm has been previously exploited in blind signal detection in conventional massive MIMO systems \cite{zhang2018blind, liu2018super}, where the channel and signal can be simultaneously estimated by factorizing the received data matrix. In contrast, our work aims to factorize the cascaded channel of the LIM-assisted massive MIMO systems, which is a completely different problem from the one considered in \cite{zhang2018blind, liu2018super}.

{\it Notations:} $\mbs{E}\{\cdot\}$, $\mbs{Var}\{\cdot\}$, $\delta(\cdot)$, $\mbs{C}$ ($\mbs{R}$), $(\cdot)^T$, $(\cdot)^H$, $(\cdot)^\ast$, and $\|\cdot\|_F$ denote the expectation operator, variance operator, Dirac delta function, space of complex (real) number, transpose, conjugate transpose, conjugate operation, and Frobenius norm, respectively (resp.). The $(i,j)$-th entry, $i$-th row, and $j$-th column of a matrix $\mbf{A}$, are denoted by $a_{i,j}$, $\mbf{a}_i^T$, and $\bsm{a}_j$, resp. We use $ \odot $ and $\mcl{CN}(\bsm{a},\bsm{C})$ to stand for the Hadamard product and the circularly-symmetric complex Gaussian distribution with mean vector $\bsm{a}$ and covariance matrix $\bsm{C}$, resp.

\section{System Model}

\begin{figure}[t]
    \centering
    \includegraphics[scale=0.9]{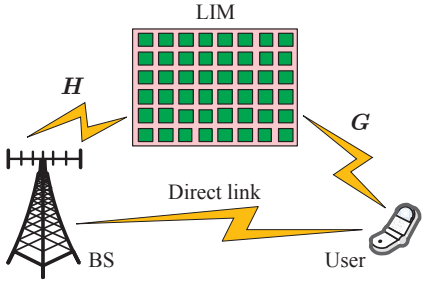}
    \vspace{-0.2cm}
    \caption{A LIM-assisted massive MIMO system.}
    \label{fig1}
    \vspace{-0.01cm}
\end{figure}

Consider a LIM-assisted massive MIMO system, as shown in Fig. \ref{fig1}, where the LIM consists of $N$ low-cost passive elements and the BS is equipped with $M$ transmit antennas to serve a number of user terminals, each equipped with $L$ receive antennas. In particular, the LIM is deployed to assist the BS in communicating with the users. Without loss of generality, we consider the communication from the BS to a reference user. The baseband equivalent channels from the BS to the LIM and from the LIM to the user are resp. denoted by $\bsm{H} \in \mbs{C}^{N \times M}$ and $\bsm{G} \in \mbs{C}^{L \times N} $. The channel component of the direct link between the BS and the user is neglected due to unfavorable propagation conditions or can be estimated (and cancelled from the model) via conventional massive MIMO channel estimation methods (see e.g., \cite{gao2015spatially, zhang2018blind, huang2018iterative}) by turning off the LIM. We assume a block-fading channel with coherence time $T$, i.e., the channel remains unchanged within each transmission block of length $T$\footnote{We assume that the channel varies from block to block. This model naturally arises when any two of the three nodes (i.e., the BS, the LIM, and the user) are in a moving state. Note that the BS and the LIM may be fixed at known positions in practice. We emphasize that even in this situation the block fading assumption may still be valid when considering millimeter wave communications in bad weather conditions (e.g., heavy rain, downpour, and monsoon) \cite{pi2011introduction}.}. Then, the received signal of the reference user can be expressed as
\begin{align}\label{received-vector}
\bsm{y}[t] =  \bsm{G} \big( \bsm{s}[t] \odot (\bsm{H} \bsm{x}[t] ) \big) + \bsm{w}[t],\,t = 1,\ldots,T
\end{align}
where $\bsm{x}[t] \in \mbs{C}^M$ and $\bsm{w}[t] \in \mbs{C}^L $ are resp. the transmitted signal and the additive noise drawn from $\mcl{CN}(\mbf{0},\sigma^2 \bsm{I})$ at time $t$ with $\sigma^2$ being the noise power. The phase shift vector $\bsm{s}[t]$ of the LIM is defined as $\bsm{s}[t] \triangleq [s_{1,t} e^{\jmath \theta_{1,t}},\ldots,$ $s_{N,t} e^{\jmath \theta_{N,t}} ]^T$ where $\jmath \triangleq \sqrt{-1}$, with $\theta_{n,t} \in (0,2\pi]$ and $s_{n,t} \in \{0,1\}$ representing the phase shift and the on/off state\footnote{The state ``off'' of a LIM reflect element means that there is only structure-mode reflection generated as if the element is a conducting object, whereas the state ``on'' means that there are both structure-mode reflection and antenna-mode reflection \cite{green1963general}. Note that the structure-mode can be absorbed into the direct link in channel modelling.} of the $n$-th LIM reflect element at time $t$, resp. By summarizing all the $T$ samples in a transmission block, the received signal can be recast as
\begin{align} \label{received-matrix}
\bsm{Y} = \bsm{G} \big( \bsm{S} \odot ( \bsm{H} \bsm{X} ) \big) + \bsm{W}
\end{align}
where $\bsm{Y} \! \triangleq \! \big[ \bsm{y}[1],\ldots,\bsm{y}[T] \big] \in \mbs{C}^{L \times T}$, $\bsm{W} \! \triangleq \! \big[\bsm{w}[1],\ldots, \bsm{w}[T] \big]$ $ \in \mbs{C}^{L \times T}$, $\bsm{S} \triangleq [\bsm{s}[1],\ldots,\bsm{s}[T]\big] \in \mbs{C}^{N \times T}$, and $\bsm{X} \triangleq \big[\bsm{x}[1],\ldots, $ $\bsm{x}[T]\big] \in \mbs{C}^{M \times T}$. In general, a smart programmable controller is built in the LIM to adaptively adjust the states and phases of the LIM based on environmental changes, which is usually referred to as reflect beamforming or passive beamforming \cite{wu2018intelligent, yan2019}.

The BS-LIM channel matrix $\bsm{H}$ and the LIM-user channel matrix $\bsm{G}$ are assumed to be rank-deficient, i.e., ${\rm rank}(\bsm{H}) < \min\{N,M\}$ and ${\rm rank}(\bsm{G}) < \min\{L,N\}$. The rank-deficiency property commonly arises in modeling massive MIMO channels under far-field and limited-scattering assumptions \cite{tse2005fundamentals}, particularly when the carrier frequency goes to millimeter wave band \cite{wang2014tens}.

\section{Problem Statement}

The objective of this paper is to estimate the cascaded channel matrices $\bsm{H}$ and $\bsm{G}$ based on the observation $\bsm{Y}$ by assuming that the pilot symbols $\bsm{X}$ and $\bsm{S}$ are known to the receiver. We note that the knowledge of both $\bsm{H}$ and $\bsm{G}$ are required in reflect beamforming \cite{wu2018intelligent}, energy-efficient design \cite{huang2018large}, as well as in simultaneous passive beamforming and information transfer \cite{yan2019}. In general, in \cite{wu2018intelligent, huang2018large, yan2019} the phase shift vectors $\{\bsm{s}[t]\}$ are set as a constant vector in each transmission block\footnote{In reflect/passive beamforming, there is no need to change the value of the phase shift vector $\bsm{s}[t]$ unless the channel changes.}, i.e.,
\begin{align}
\bsm{s}[t] = \bsm{s}, t = 1,\ldots,T. \notag
\end{align}
Then, the signal model in \eqref{received-matrix} reduces to
\begin{align} \label{received-constant}
\bsm{Y} = \bsm{G} {\rm diag}\{\bsm{s}\} \bsm{H} \bsm{X} + \bsm{W}.
\end{align}
Clearly, we have
\begin{align} \label{permutation1}
\bsm{G} {\rm diag}\{\bsm{s}\} \bsm{H} = \bsm{G}' {\rm diag} \{\bsm{s}\} \bsm{H}'
\end{align}
where $ \bsm{G}' \triangleq \bsm{G} \bsm{\Phi}$ and $\bsm{H}' \triangleq \bsm{\Phi}^{-1} \bsm{H}$ are effective channels in the sense that \eqref{permutation1} holds, and $\bsm{\Phi} \in \mbs{C}^{N \times N}$ is an invertible diagonal matrix. This implies that the optimization of $\bsm{s}$ for reflect beamforming based on $\bsm{G}$ and $\bsm{H}$ yields the same result as that based on $\bsm{G}'$ and $\bsm{H}'$. Therefore, the optimization of $\bsm{s}$ for  reflect beamforming in \eqref{received-constant} requires the knowledge of $\bsm{G}$ and $\bsm{H}$ up to the ambiguity of a full-rank diagonal matrix $\bsm{\Phi}$. In other words, we need to estimate the cascaded channel $\bsm{G}$ and $\bsm{H}$ up to the ambiguity in \eqref{permutation1} by an appropriate design of the training signals $\{\bsm{s}[t]\}$ and $\{\bsm{x}[t]\}$.

\section{Proposed Channel Estimation Algorithm}

We now address the cascaded channel estimation issue of the LIM-aided massive MIMO systems described in the preceding section. Notice that setting $\bsm{s}[t]$ as a constant in a transmission block as in \eqref{received-constant} does not work for the cascaded channel estimation. In that case, for given $\bsm{S} = {\rm diag}\{\bsm{s}\}$ and $\bsm{X}$, the estimates of both $\bsm{G}$ and $\bsm{H}$ from $\bsm{Y}$ in \eqref{received-constant} can be interpreted as an affine matrix factorization problem \cite{liu2018super}. The condition that $\bsm{G}$ and $\bsm{H}$ are both rank-deficient is not enough to guarantee the successful recovery of $\bsm{G}$ and $\bsm{H}$ from $\bsm{Y}$. As such, we propose to employ a varying $\bsm{s}[t]$ in the design of training signals, as detailed below.

The received signal in \eqref{received-matrix} can be recast as a bilinear model
\begin{align} \label{Bilinear-sparse}
\bsm{Y} = \bsm{G} \bsm{Z} + \bsm{W}
\end{align}
where $\bsm{Z}  = \bsm{S} \odot ( \bsm{H} \bsm{X} )$. Our proposed method is to estimate $\bsm{G}$ and $\bsm{Z}$ first and then $\bsm{H}$ based on the estimate of $\bsm{Z}$. To facilitate the estimation of the cascaded channels $\bsm{G}$ and $\bsm{H}$, we describe the constructions of pilot symbols $\bsm{S}$ and $\bsm{X}$ as follows. For the LIM pilot $\bsm{S}$, we generate $\{s_{t,n}\}$ independently from a bernoulli distribution ${\rm Bernoulli}(\lambda)$ with $\lambda$ being the probability of taking the value of $1$ such that partial samples from $\bsm{Z}$ are available. The phase shifts $\{\theta_{n,t}\}$ can be set to any value within $(0, 2\pi]$ because  they would not affect the on/off states of the elements in the LIM. The transmitted pilot $\bsm{X}$ is required to be a full-rank matrix, i.e., ${\rm rank}(\bsm{X}) = \max (M,T)$, which will become clear in Section \ref{MC}.

From the construction of $\bsm{S}$, we see that $\bsm{Z}$ in \eqref{Bilinear-sparse} is a sparse matrix with a large number of zero elements when the sparsity level (sampling rate) $\lambda$ becomes small.
Thus, we can recover $\bsm{G}$ and $\bsm{Z}$ from $\bsm{Y}$ via the available bilinear sparse matrix factorization techniques, such as the BiG-AMP \cite{parker2014bilinear}. Note that the sparse matrix factorization suffers from a diagonal ambiguity when the support information of $\bsm{Z}$ is given \cite{zhang2018blind}. That is, if $(\hat{\bsm{G}},\hat{\bsm{Z}})$ is a solution to the bilinear factorization problem, then alternative $(\hat{\bsm{G}} {\bsm{\Phi}}, {\bsm{\Phi}}^{-1} \hat{\bsm{Z}})$ is also a valid solution, where $\bsm{\Phi}$ is an invertible diagonal matrix. Recall that there is no need to eliminate the diagonal ambiguity $\bsm{\Phi}$ based on \eqref{permutation1} and the discussions therein. Then, based on the estimated $\hat{\bsm{Z}}$ and the known $\bsm{S}$, the next step is to retrieve $\bsm{H}$ by using matrix completion to fill the missing entries of $\hat{\bsm{Z}}$, together with the fact that $\bsm{H}$ is rank-deficient. Such a combined two stage channel estimation algorithm is referred to as the joint bilinear factorization and matrix completion (JBF-MC) algorithm which is presented in Algorithm \ref{algorithm1}. The details of the JBF-MC algorithm are elaborated in the following subsections.

\subsection{Sparse Matrix Factorization}

Based on the observation $\bsm{Y}$, we use the BiG-AMP algorithm \cite{parker2014bilinear} to approximately calculate the minimum mean-squared error estimates of $\bsm{G}$ and $\bsm{Z}$, i.e., the means of the marginal posteriors $\{p(g_{l,n} | \bsm{Y})\}$ and $\{p(z_{n,t} | \bsm{Y})\}$. Specifically, the BiG-AMP implements the sum-product loopy belief propagation over a factor graph induced by the the posterior distribution
\begin{align} \label{MAP}
p(\bsm{G},\bsm{Z}|\bsm{Y}) \varpropto p(\bsm{Y}|\bsm{B}) p(\bsm{G}) p (\bsm{Z})
\end{align}
where $\bsm{B} = \bsm{G} \bsm{Z}$ and the likelihood $p(\bsm{Y}|\bsm{B})$ is given as
\begin{align}
p(\bsm{Y}|\bsm{B}) & = \prod_{l=1}^L \prod_{t=1}^T p(y_{l,t}|b_{l,t}) \notag \\
& = \prod_{l=1}^L \prod_{t=1}^T \exp\left( -{|y_{l,t} - b_{l,t} |^2}/{\sigma^2}  \right).
\end{align}
We choose independent Gaussian priors for $\bsm{G}$ and independent Bernoulli-Gaussian priors for $\bsm{Z}$, i.e.,
\begin{align}
p(\bsm{G}) & =  \prod_{l=1}^L \prod_{n=1}^N p(g_{l,n}) = \prod_{l=1}^L \prod_{n=1}^N \mcl{CN}(g_{l,n};0,\nu_g) \label{prior-G}\\
p(\bsm{Z}) & = \prod_{n=1}^N \prod_{t=1}^T p(z_{n,t})  = \prod_{n=1}^N \prod_{t=1}^T \big[ s_{n,t}\,\mcl{CN}(z_{n,t};0, \nu_z) \notag \\
 &~~~~~~~~~~~~~~~~~~~~~~~~~~~~~~~~~~~~~~~~~~~~~~~\, + (1-s_{n,t}) \delta(z_{n,t}) \big] \label{prior-Z}
\end{align}
where $\nu_g$ and $\nu_z$ are resp. the average variances of the LIM-user channel matrix $\bsm{G}$ and the non-zero elements of the sparse matrix $\bsm{Z}$, resp; the on/off state $s_{n,t} \in \{0,1\}$ at time $t$ for the $n$-th reflect element of the LIM is known by the receiver because $\bsm{S}$ is a pilot signal.

\begin{algorithm}[t]
\xiaowuhao
\vspace{0.2mm}
\caption{\textbf{\!:} \vspace{0.15mm} JBF-MC algorithm}
\label{algorithm1}
\begin{algorithmic}[1]
\REQUIRE $\!\bsm{Y}$, $\bsm{S}$, $\bsm{X}$, prior distributions $p(\bsm{G})$ and $p(\bsm{Z})$ \\
\vspace{0.01cm}
\hspace{-0.5cm} \% sparse matrix factorization via BiG-AMP
\STATE Initialization: $\forall l,n,t$: generate $g_{l,n}$ from $p(g_{l,n})$, $v^g_{l,n}(1) = \nu_g$, $\hat{z}_{n,t}(1) = \mbs{E}(z_{n,t}) $, $v^z_{n,t}(1) = \lambda \nu_z $, and $\hat{u}_{l,t}(1) = 0$ \\
\STATE \textbf{for} $i=1,\ldots,I_{\rm max}$~~\% outer iteration \\
\vspace{-0.05cm}
\STATE~~~\textbf{for} $j=1,\ldots,J_{\rm max}$~~\% inner iteration \\
\STATE~~~$\forall l,t$: $\bar{v}_{l,t}^p(i) =\sum_{n=1}^{N}|\hat{g}_{l,n}(i)|^2 v^z_{n,t}(i)
+ v^g_{l,n}(i) |\hat{z}_{n,t}(i)|^2$ \\
\STATE~~~$\forall l,t$: $\bar{p}_{l,t}(i)=\sum_{n=1}^{N}\hat{g}_{l,n}(i)\hat{z}_{n,t}(i)$ \\
\STATE~~~$\forall l,t$: $v_{l,t}^p(i)=\bar{v}_{l,t}^p(i)+\sum_{n=1}^{N}v_{l,n}^g(i)v_{n,t}^z(i)$ \\
\STATE~~~$\forall l,t$: $\hat{p}_{l,t}(i) = \bar{p}_{l,t}(i)-\hat{u}_{l,t}(i-1) \bar{v}_{l,t}^p(i)$\\
\STATE~~~$\forall l,t$: $v_{l,t}^b (i) = {\sigma^2  v_{l,t}^p(i)}/\big[v_{l,t}^p(i)+\sigma^2 \big]$\\
\STATE~~~$\forall l,t$: $\hat{b}_{l,t}(i)=v_{l,t}^p(i) [y_{l,t}-\hat{p}_{l,t}(i)]/[v_{l,t}^p(i)
+\sigma^2 ]+\hat{p}_{l,t}(i)$ \\
\STATE~~~$\forall l,t$: $v_{l,t}^u(i)=\big[ 1-v_{l,t}^z(l)/v_{l,t}^p(i) \big]/v_{l,t}^p(i)$ \\
\STATE~~~$\forall l,t$: $\hat{u}_{l,t}(i)=\big[\hat{b}_{l,t}(i)-\hat{p}_{l,t}(i)\big]/v_{l,t}^p (i)$ \\
\STATE~~~$\forall l,n$: $v_{l,n}^q(i)=\big[ \sum_{t=1}^{T}|\hat{z}_{n,t}(i)|^2 v_{l,t}^u (i) \big]^{-1}$ \\
\STATE~~~$\forall l,n$: $\hat{q}_{l,n}(i) = \hat{g}_{l,n}(i) \big[1- v_{l,n}^q(i) \sum_{t=1}^{T}v_{n,t}^{z}(i) v_{l,t}^u(i) \big]$ \\
$~~~~~~~~~~~~~~~~~~~~~~~~~~~~~ +v_{l,n}^q(i)\sum_{t=1}^{T} \hat{z}^\ast_{n,t}(i) \hat{u}_{l,t}(i)$ \\
\STATE~~~$\forall n,t$: $v_{n,t}^r(i)=\big[\sum_{l=1}^{L}|\hat{g}_{l,n}(i)|^2 v_{l,t}^u(i)\big]^{-1}$ \\
\STATE~~~$\forall n,t$: $\hat{r}_{n,t}(i)=\hat{z}_{n,t}(i) \big(1- v^r_{n,t}(i) \sum_{l=1}^{L} v_{l,n}^g(i) v_{l,t}^u (i) \big)$\\
$~~~~~~~~~~~~~~~~~~~~~~~~~~~~~+v_{n,t}^r(i)\sum_{l=1}^{L}\hat{g}^\ast_{l,n}(i)  \hat{u}_{l,t}(i)$ \\
\STATE~~~$\forall l,n$: $\hat{g}_{l,n}(i+1)= \mbs{E} \big\{ g_{l,n}|\hat{q}_{l,n}(i), v_{l,n}^q(i) \big\}$ \\
\STATE~~~$\forall l,n$: $v_{l,n}^g (i+1) = \mbs{Var}\big\{ g_{l,n}|\hat{q}_{l,n}(i), v_{l,n}^q(i) \big\}$ \\
\STATE~~~$\forall n,t$: $\hat{z}_{n,t}(i+1)=\mbs{E} \big\{ z_{n,t}|\hat{r}_{n,t}(i), v_{n,t}^r(i)\big\}$ \\
\STATE~~~$\forall n,t$:
$v_{n,t}^z(i+1)= \mbs{Var} \big\{ z_{n,t}|\hat{r}_{n,t}(i), v_{n,t}^r(i)\big\} $ \\
\STATE~~~\textbf{if} {\it a certain stopping criterion is met}, \textbf{stop}
\vspace{-0.05cm}
\STATE~~~\textbf{end for} \\
\vspace{-0.1cm}
\STATE~~~$\forall l, n, t, \hat{g}_{l, n}(i)=\hat{g}_{l, n}(i+1), \nu_{m, n}^g(i)=\nu_{l, n}^{g}(i+1)$, \\ ~~~$\hat{z}_{n, t}(i)=\hat{z}_{n, t}(1), \nu_{n, t}^{z}(i)= v^z_{n,t}(1) $ \\
\vspace{-0.1cm}
\STATE\,\textbf{end for} \\
\STATE $\hat{\bsm{G}} \gets \hat{\bsm{G}}(i+1)$, $\hat{\bsm{Z}} \gets \hat{\bsm{Z}}(i+1)$ \\
\vspace{0.4mm}
\hspace{-0.6cm} \% matrix completion via RGrad
\STATE  Initialization: $\bsm{A}(0) = \bsm{0}$
\STATE \textbf{for} $k=1,\ldots,K_{\rm max}$
\STATE~~~$\bsm{Q}(k) = \bsm{S}^\ast \odot \big(\hat{\bsm{Z}} - \bsm{A}(k)\big)$ \\
\STATE~~~$\alpha(k) = \frac{\| \mcl{P}_{\mcl{S}(k)}(\bsm{Q}(k)) \|_F^2}{\|\bsm{S}^\ast \odot (\mcl{P}_{\mcl{S}(k)}(\bsm{Q}(k)) ) \|_F^2}$ \\
\STATE~~~$\bsm{W}(k) = \bsm{A}(k) + \alpha_{k} \mathcal{P}_{\mathcal{S}(k)} \left(\bsm{Q}(k) \right)$ \\
\STATE~~~$\bsm{A}(k+1)=\mathcal{H}_{r}\left(\bsm{W}(k)\right)$ \\
\STATE~~~\textbf{if} {\it a certain stopping criterion is met}, \textbf{stop}
\STATE\textbf{end for} \\
\STATE $\hat{\bsm{H}} \gets \hat{\bsm{A}} \bsm{X}^\dagger$ with $\hat{\bsm{A}} = \bsm{A}(k+1)$ \\
\hspace{-0.55cm}\textbf{\,Output:} $\hat{\bsm{G}}$ and $\hat{\bsm{H}}$
\end{algorithmic}
\end{algorithm}

To achieve computational efficiency, the BiG-AMP leverages the central limit theorem and the quadratic approximation for the involved messages. Details of the BiG-AMP algorithm can be found from Lines 1 to 24 of Algorithm 1. At the $i$-th iteration $i$ of the BiG-AMP, the means and variances of $\{g_{l,n}\}$ and $\{z_{n,t}\}$ from Lines 16 to 19 are resp. calculated with respect to the approximate marginal posterior distributions:
\begin{align}
\hat{p}^{(i)}(g_{l,n}) & \propto p(g_{l,n})\,\mcl{CN} \big( g_{l,n}; \hat{q}_{l,n}(i),  v^q_{l,n} (i) \big) \\
\hat{p}^{(i)}(z_{n,t}) & \propto p(z_{n,t})\, \mcl{CN} \big( z_{n,t}; \hat{r}_{n,t}(i),  v^r_{n,t} (i) \big)
\end{align}
where $p(g_{l,n})$ and $p(z_{n,t})$ are the prior distributions defined in \eqref{prior-G} and \eqref{prior-Z}, resp.

It is worth noting that the K-SVD (K-means singular value decomposition) \cite{aharon2006k} and the SPAMS (SPArse Modeling Software) \cite{mairal2010online} can be exploited to solve the bilinear sparse matrix factorization problem \eqref{Bilinear-sparse}. However, these approaches  perform much worse than the BiG-AMP algorithm, as will be seen from the numerical results presented in Section \ref{section-v}. Additionally, notice that the likelihood $p(\bsm{Y} | \bsm{B})$ and the priors $p(\bsm{G})$ and $p(\bsm{H})$ are characterized by the parameters $\sigma^2$, $\nu_g$, and $\nu_z$, resp. These parameters can be estimated by leveraging the standard Expectation-Maximization methodology \cite{parker2014bilinear, huang2018iterative}. In this work, we do not elaborate on how to estimate these parameters due to space limitation.

\vspace{-0.05cm}

\subsection{Matrix Completion} \label{MC}

To facilitate the estimate of $\bsm{H}$, we now recover the missing entries of $\hat{\bsm{Z}}$ by using the rank-deficient property of the original $\bsm{H}$. We employ the Riemannian gradient (RGrad) algorithm \cite{wei2016guarantees}
to solve the matrix completion problem, which is summarized from Lines 25 to 32 of Algorithm 1. The RGrad algorithm is to solve the following matrix  completion problem:
\begin{align}
\min_{\bsm{A}}~\frac{1}{2} \big\| \bsm{S}^\ast \odot ( {\bsm{A}} -  \hat{\bsm{Z}} ) \big\|_F^2~~{\rm subject~to}~ {\rm rank}({\bsm{A}}) = r.
\end{align}
In Lines 28 and 29 of the JBF-MC algorithm, $\mcl{P}_{\mcl{S}(k)}(\cdot)$ stands for the projection operation to the left singular vector subspace (denoted by $\mcl{S}(k)$) of the current estimate $\bsm{A}(k)$, corresponding to the first $r$ eigenvalues of $\bsm{A}(k)$. In Line 30 of Algorithm 1, $\mathcal{H}_{r}\left(\bsm{W}\right)$ is the hard-thresholding operator for the best rank-$r$ approximation of the associated SVD, i.e.,
\begin{align}
\mathcal{H}_{r}\left(\bsm{W}\right) \triangleq \bsm{U} \bsm{\Sigma}_r \bsm{V}^H,\,\Sigma_r(i,i) =  \begin{cases}
 \Sigma(i,i), \!\!\!\!& i \leq r \\
 0, \!\!\!\! & i > r
 \end{cases}
\end{align}
where $\bsm{W} \triangleq \bsm{U} \bsm{\Sigma} \bsm{V}^H$ is the SVD of $\bsm{W}$. Finally, the estimate of the channel matrix $\bsm{H}$ can be computed as
\begin{align} \label{H-estimate}
\hat{\bsm{H}} = \hat{\bsm{A}} \bsm{X}^\dagger
\end{align}
where $\bsm{X}^\dagger = (\bsm{X} \bsm{X}^H)^{-1} \bsm{X} $ is the Moore-Penrose inverse and $\hat{\bsm{A}} $ is output of the RGrad algorithm. Here, we assume that the pilot length $T$ is no less than the number of transmit antennas $M$ and ${\rm rank(\bsm{X})} = M$, so as to ensure the existence of $\bsm{X}^\dagger$.
\vspace{-0.05cm}

\subsection{Computational Complexity}

We now offer a brief discussion on the computational complexity of the proposed JBF-MC algorithm. Note that the total computational complexity of the JBF-MC consists of implementations of the BiG-AMP for matrix factorization and the RGrad for matrix completion and the computation involved in \eqref{H-estimate}. We thus sketch the respective complexity as follows. First, the complexity of the BiG-AMP is dominated by basic matrix multiplications in Lines 4--6 and Lines 12--15 of Algorithm 1, requiring $\mcl{O}(LNT)$ flops per iteration. Consequently, the complexity of the BiG-AMP is at most $I_{\rm max} J_{\rm max} \mcl{O}(LNT)$ flops, where $I_{\rm max}$ and $J_{\rm max}$ are resp. the maximum numbers of the outer and inner iterations of the BiG-AMP. Second, the complexity of the RGrad is dominated by the calculations in Lines 27 and 29, requiring $\mcl{O}(rNT)$ and $\mcl{O}(r^3)$ flops, resp. Thus, the total cost of the RGrad is at most $K_{\rm max} \mcl{O}(rNT)$ flops by noting that $r < \min \{N,T\}$, where $K_{\rm max}$ is the maximum number of iterations of the RGrad. Finally, the computation of \eqref{H-estimate} in Line 33 requires $\mcl{O}(M^2T) + \mcl{O}(M^3) + \mcl{O}(MNT)$ flops.

\section{Simulation Results} \label{section-v}

We now carry out numerical experiments to corroborate the effectiveness of the proposed JBF-MC algorithm for the cascaded channel estimation of the LIM-assisted massive MIMO systems. We assume that the antenna elements form a half-wavelength uniform linear array (ULA) configuration at the BS, the LIM and the receiver side. Following the superposition principle of different paths in the prorogation environment \cite{zhang2018blind, huang2018iterative}, the baseband BS-LIM and LIM-user channel matrices $\bsm{H}$ and $\bsm{G}$ are resp. modeled as
\begin{align}
\bsm{H} =& \sum_{k=1}^{K_h} \alpha_k\,\bsm{a}_L(\vartheta_{k}) \bsm{a}_T^H (\omega_{k}),\,
\bsm{G} =& \sum_{k=1}^{K_g} \beta_k\,\bsm{a}_R(\psi_{k}) \bsm{a}_L^H (\vartheta_{k}) \notag
\end{align}
where $K_h$ and $K_g$ stand for the number of propagation paths of radio signals in the BS-LIM channel and the LIM-user channel, resp.; $\bsm{a}_L(\vartheta_{k}) \in \mbs{C}^N$, $\bsm{a}_T (\omega_{k}) \in \mbs{C}^M$, and $\bsm{a}_R(\psi_{k}) \in \mbs{C}^L$ are the steering vectors of the ULA at the LIM, the BS, and the receiver side, resp. In each trial, the angular parameters $\psi_r$, $\vartheta_r$, and $\omega_{k}$ independently follow from the uniform distribution within $(0,1]$, and the path gain coefficients $\{\alpha_k\}$ and $\{\beta_k\}$ are independently drawn from $\mcl{CN}(0,1)$. The number of paths in the channel matrix $\bsm{H}$, i.e., $K_h$, is set to be enough small such that it has a low-rank structure to facilitate its estimation in the matrix completion stage. In addition, our empirical experiments suggest that choosing $K_g$ to be a positive integer more than half of the receive antenna number $L$ or the passive antenna number $N$ in the LIM can lead to a better performance. The sampling matrix $\bsm{S}$, i.e., the on/off states of the elements in the LIM, are set to be a random 0-1 matrix with fixed zero phase.

The pilot symbols in $\bsm{X}$ are generated from $\mcl{CN}(0,1)$ and the signal-to-noise ratio (SNR) is defined as $10\log_{10}(1/\sigma^2)$ dB. The estimation performance is evaluated in terms of the normalized mean-square-error (NMSE). All the simulation results are obtained by averaging $200$ independent trials. Notice that the outputs $\hat{\bsm{G}}$ and $\hat{\bsm{H}}$ of the JBF-MC algorithm still contain diagonal   ambiguities. These ambiguities are eliminated based on the true values of $\bsm{G}$ and $\bsm{H}$ in the calculation of the NMSEs. In the sparse matrix factorization stage, to benchmark our JBF-MC algorithm for the estimation of $\bsm{G}$, we adopt the K-SVD \cite{aharon2006k} and the SPAMS \cite{mairal2010online} for comparison. In all simulations, we set $N=70$ , $M=L=64$, $K_h = 4$, and $K_g = r = 35$. For comparison in the matrix completion stage, we also adopt the IHT (iterative hard thresholding) and the IST (iterative soft thresholding) \cite{cai2010singular} to estimate the BS-LIM channel $\bsm{H}$.

\begin{figure}[t]
    \centering
    \includegraphics[scale=0.32]{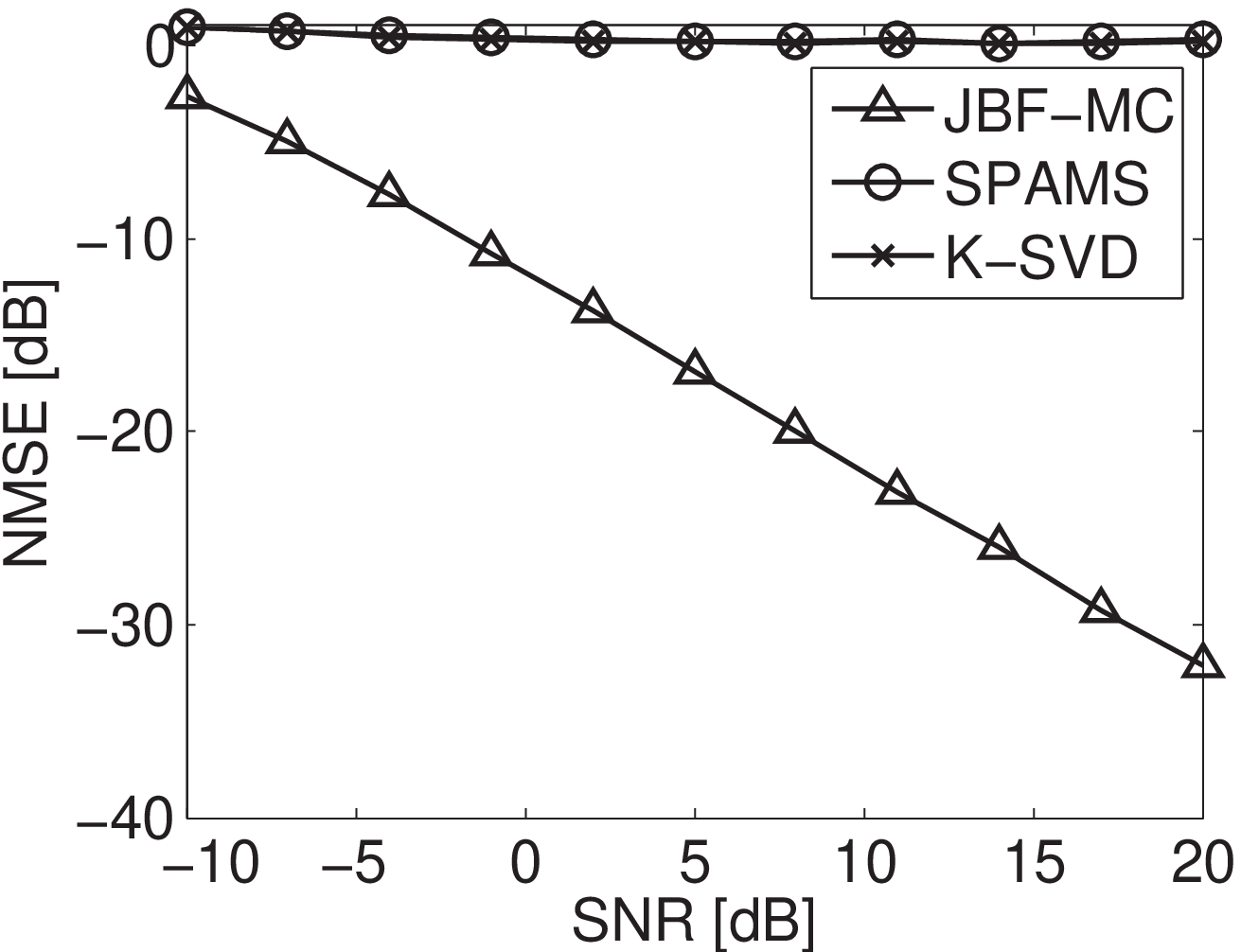}\,\, \includegraphics[scale=0.32]{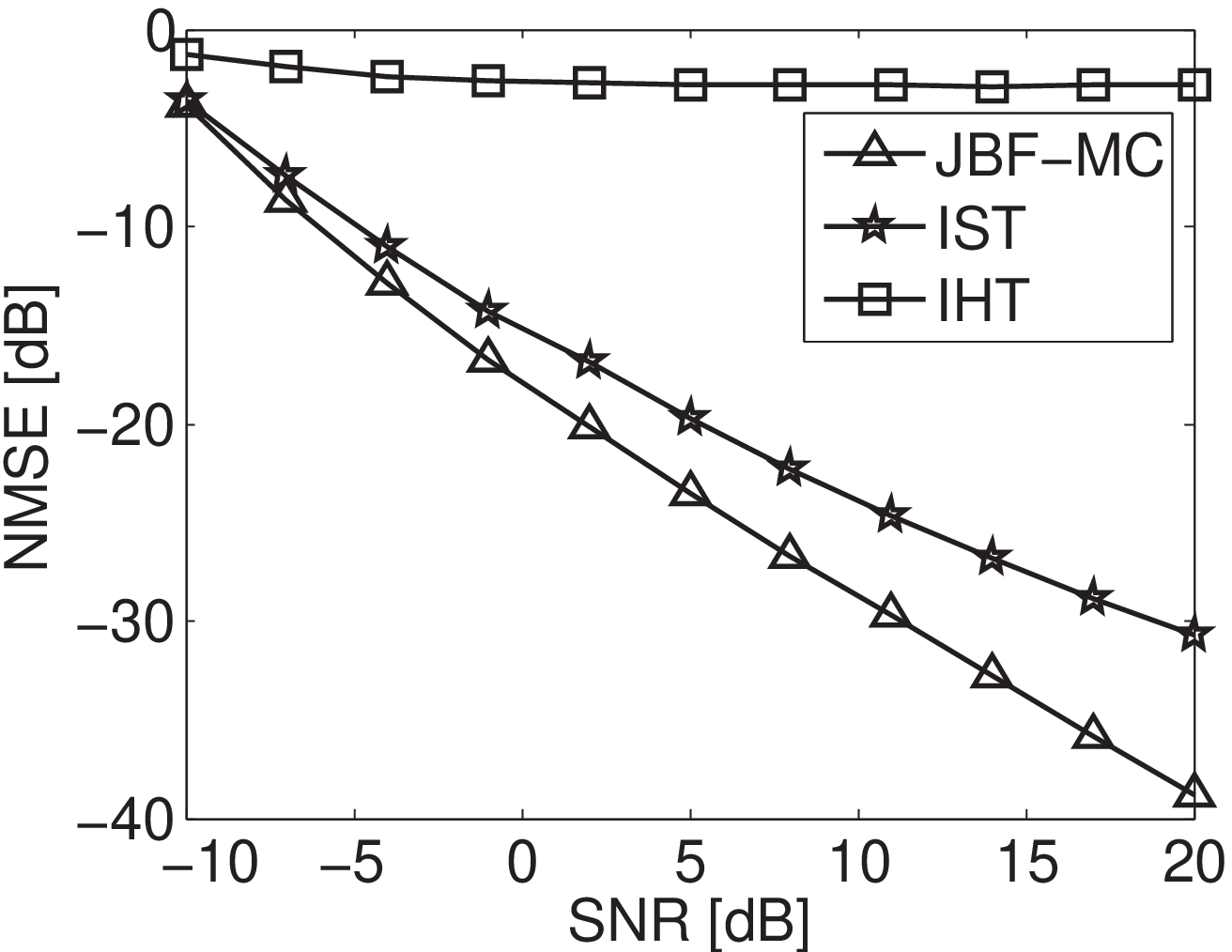} \\
    \smallskip \smallskip
    \includegraphics[scale=0.32]{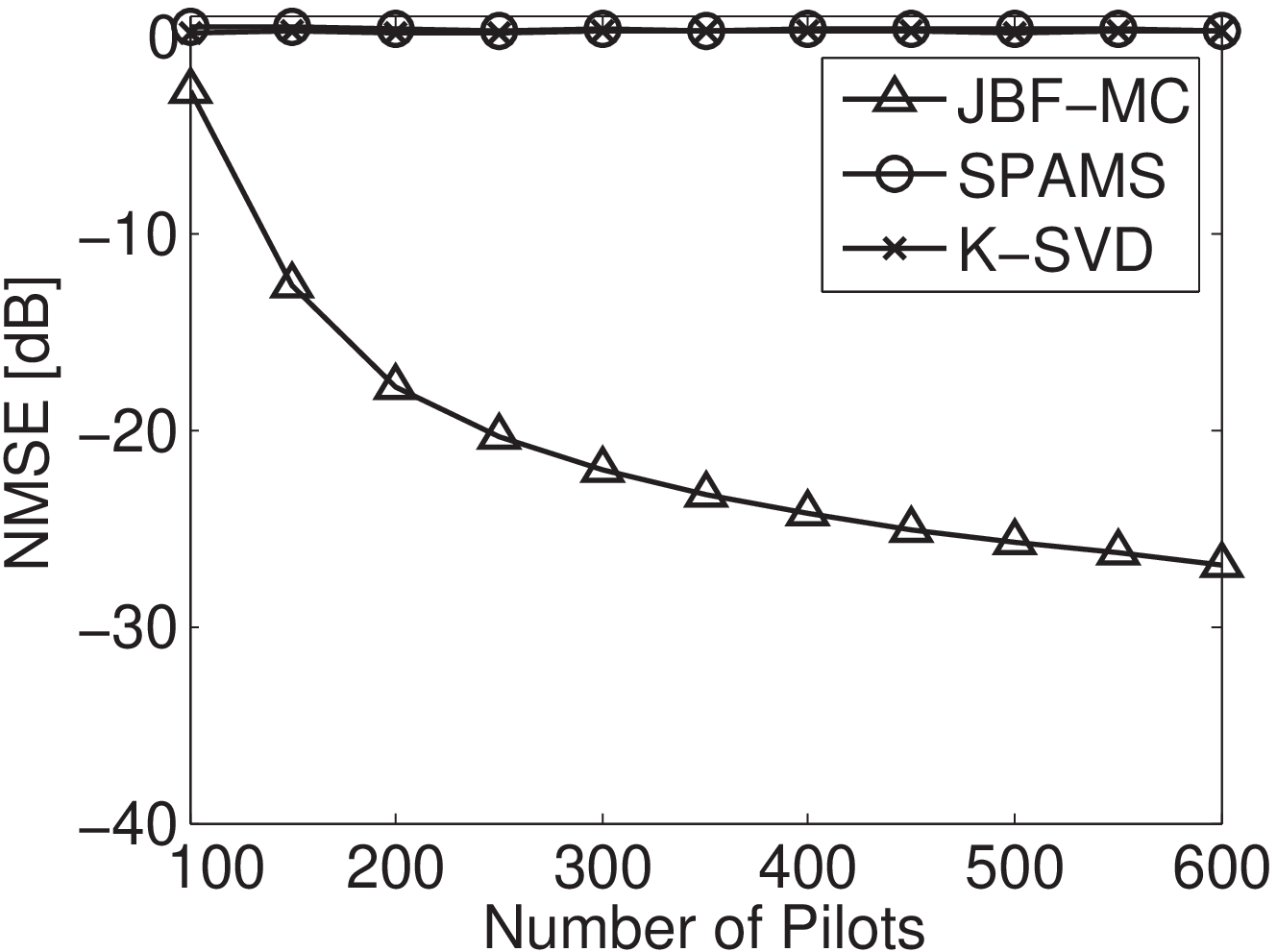}\,\, \includegraphics[scale=0.32]{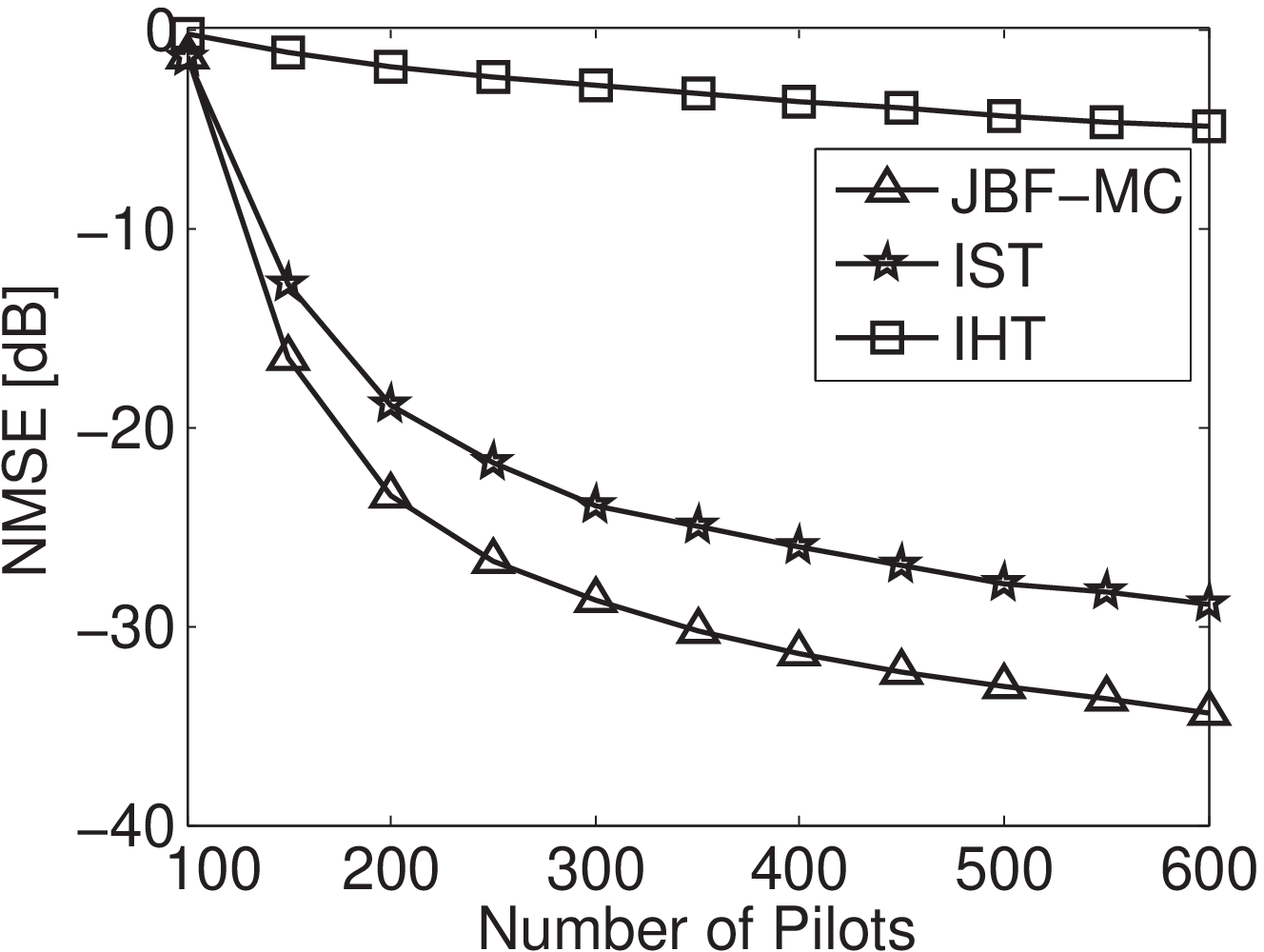}\\
    \vspace{-1mm}
    \caption{NMSEs of $\bsm{G}$ (left subplots) and $\bsm{H}$ (right subplots) versus the SNR and the number of pilots with $N=70$,   $M=L=64$, and $\lambda = 0.2$.}
    \label{fig2}
    \vspace{1mm}
\end{figure}

\begin{figure}[t]
    \centering
    \includegraphics[scale=0.32]{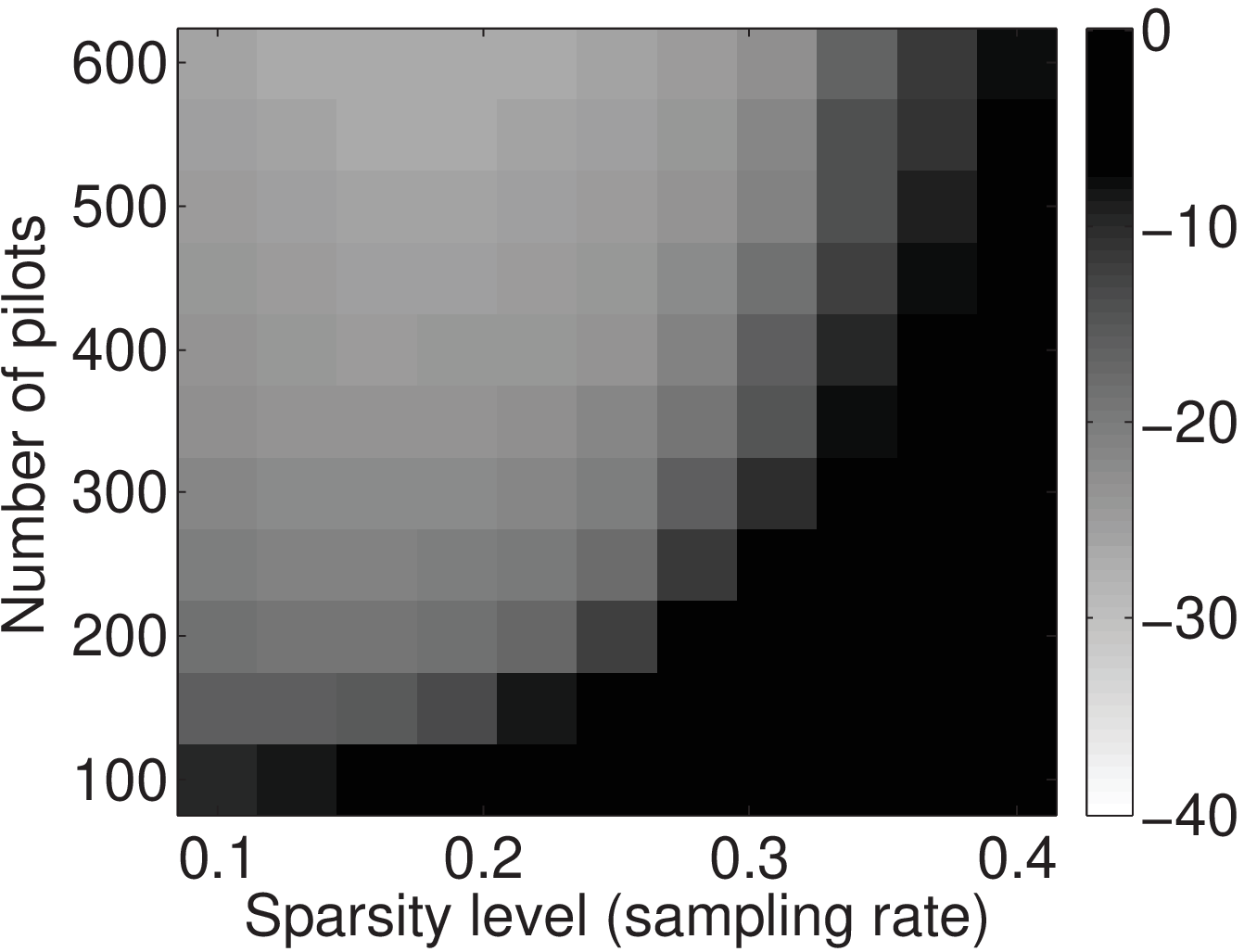}\,
    \includegraphics[scale=0.32]{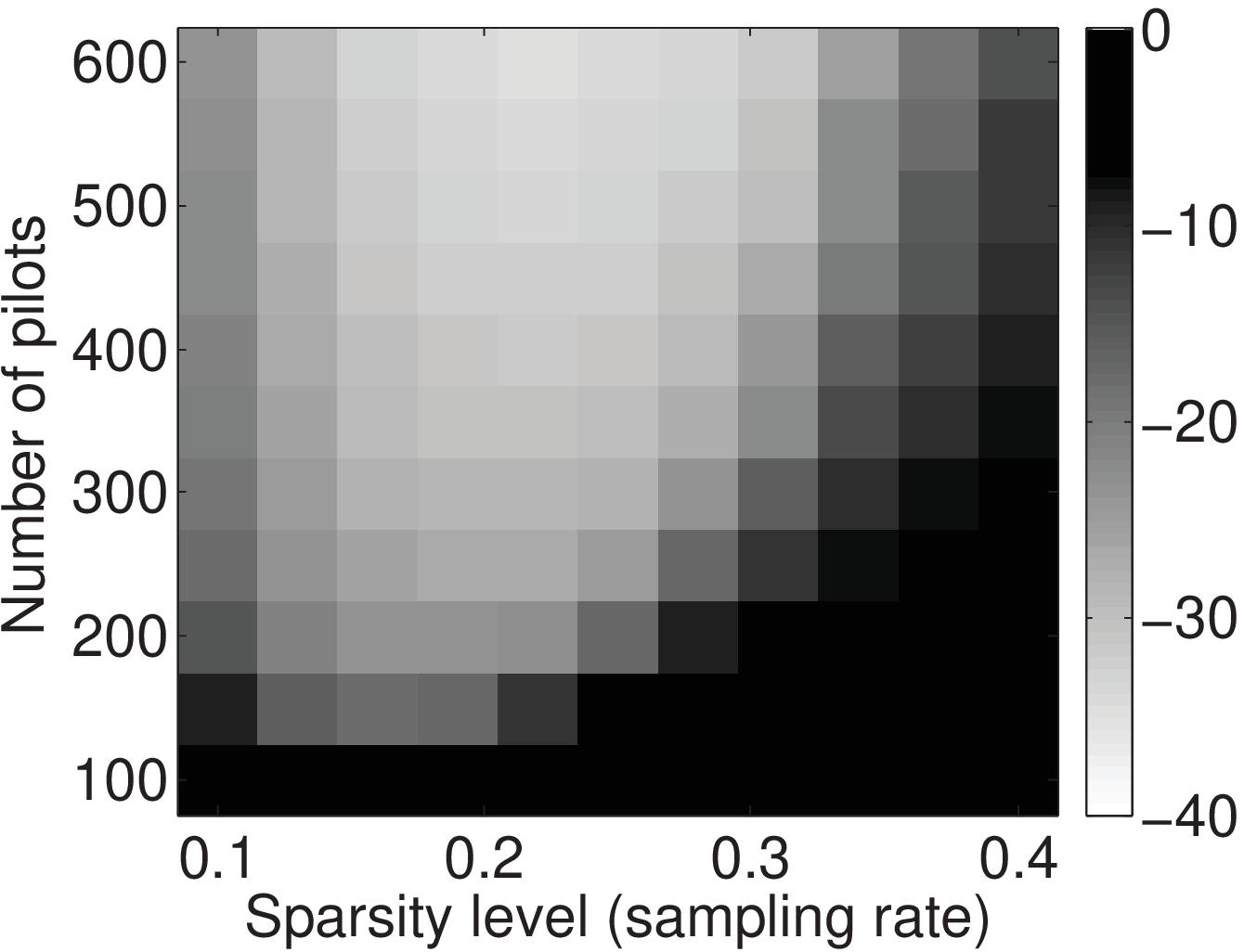}\\ \vspace{1.9mm}
    \includegraphics[scale=0.32]{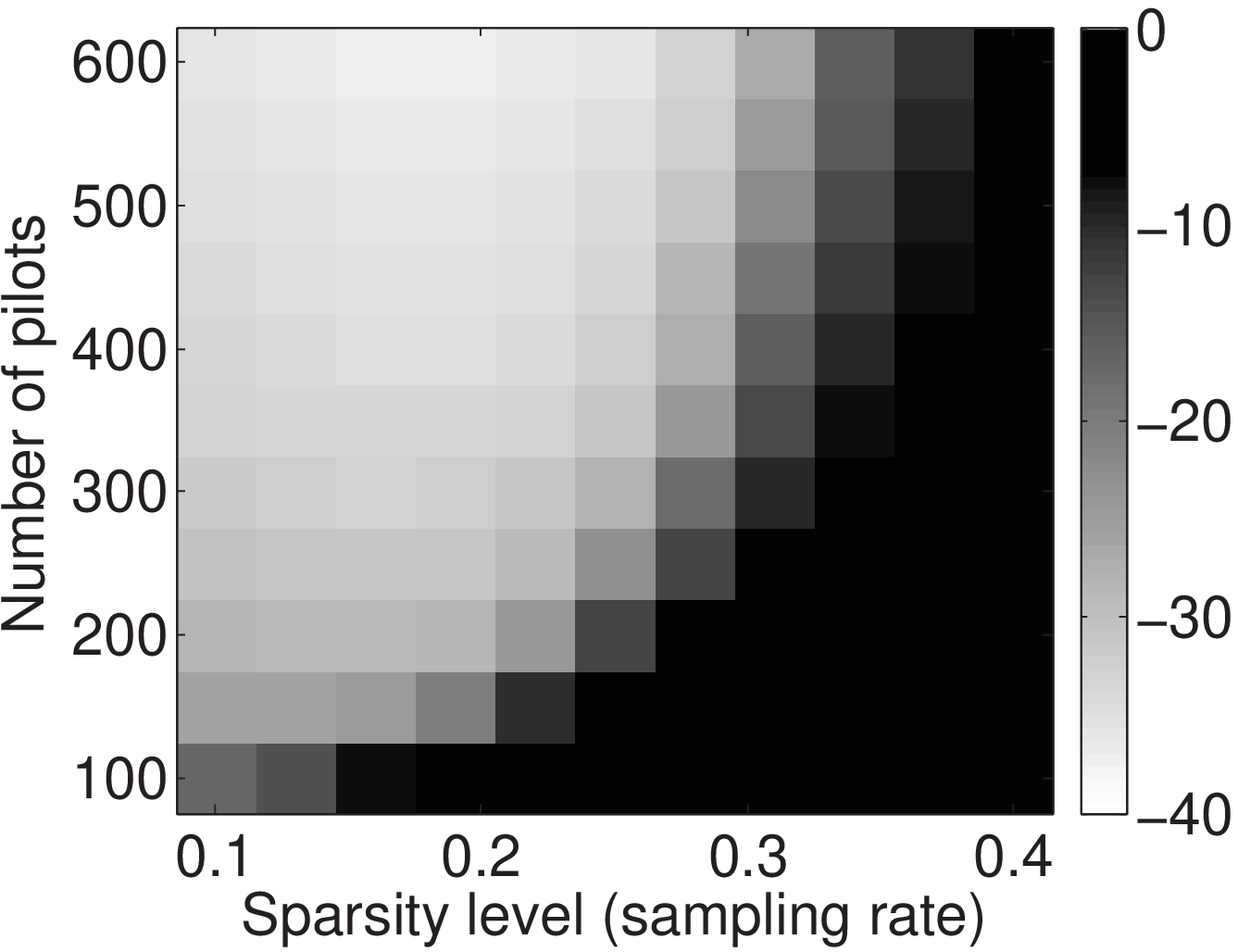}\,
    \includegraphics[scale=0.32]{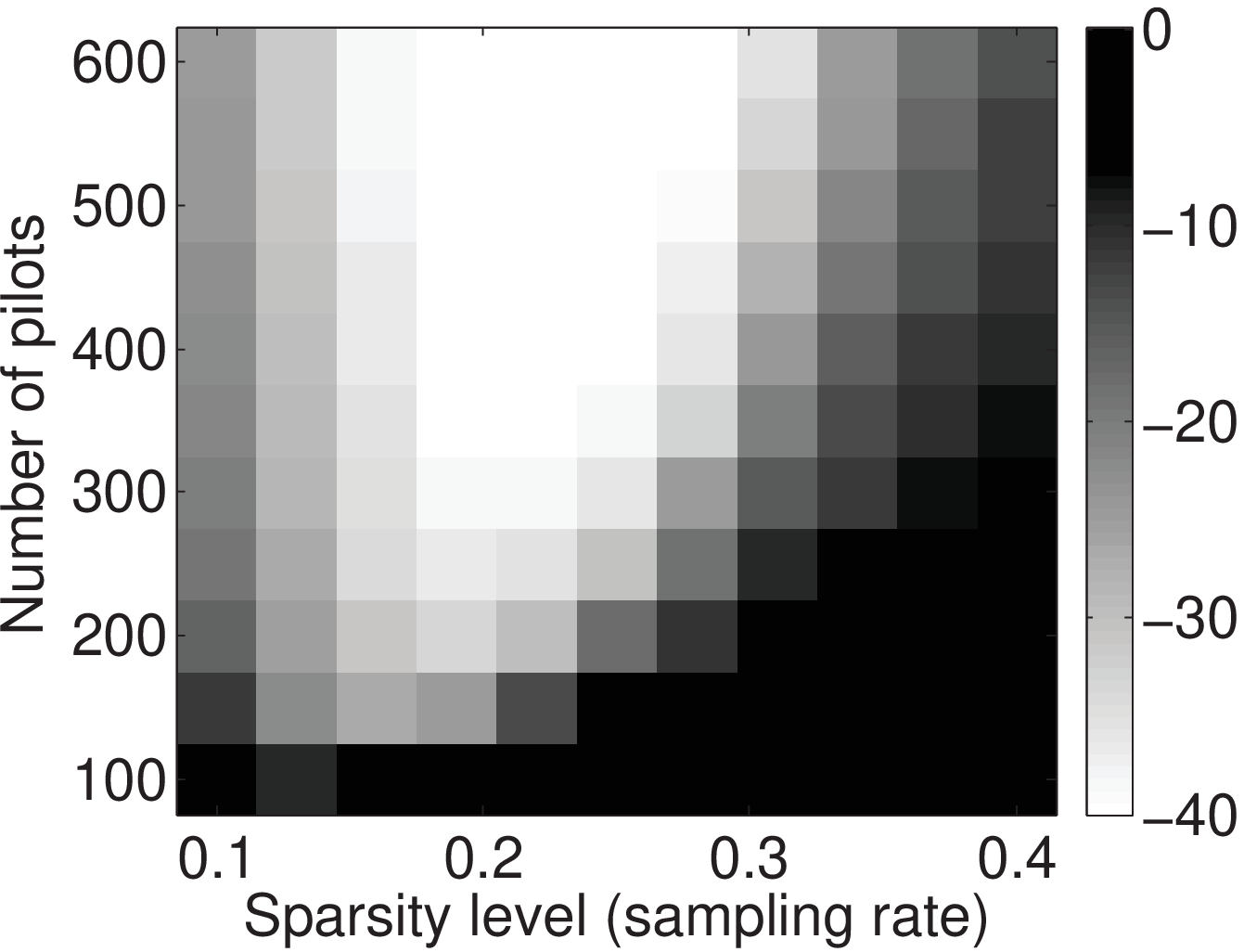}
    \vspace{-5mm}
    \caption{Phase transitions (in NMSE) of $\bsm{G}$ (left subplots) and $\bsm{H}$ (right subplots) versus the sparsity level and the number of pilots. The upper subplots are for $\text{SNR} = 10\,{\rm dB}$ and the lower subplots for $\text{SNR} = 20\,{\rm dB}$.}
    \label{fig3}
    \vspace{0mm}
\end{figure}

The NMSEs versus the SNR (with $T=300$) and the number of pilots (with  $\text{SNR} = 10\,\text{dB}$) are depicted in Fig. \ref{fig2} under the sparsity level (sampling rate) $\lambda = 0.2$. It is observed that the proposed JBF-MC algorithm has a significant performance gain over the baseline methods, especially for the estimation of $\bsm{G}$. We also observe that the NMSE of $\bsm{H}$ is smaller than that of $\bsm{G}$. This is because $\bsm{Z}$ a sparse matrix with much smaller number of variables to be estimated. Fig. \ref{fig3} shows the phase transitions of the NMSEs of the cascaded channel estimation versus the sparsity level and the number of pilots. It is seen from Fig. \ref{fig3} that there is a tradeoff for the sampling rate (sparsity level) requirement, too small of which makes the matrix completion fail but too large of which makes the sparse matrix factorization fail. This is because when a larger number of samples (i.e., large $\lambda$) are available for matrix completion, the performance of the BiG-AMP algorithm becomes more worse as an increasing number of random variables are needed to be estimated in the sparse matrix factorization stage.

\section{Conclusions}
In this work, we considered the cascaded channel estimation for the LIM-assisted massive MIMO systems. We introduced a general framework for this problem by leveraging a combined bilinear spare matrix factorization and matrix completion, and presented a two-stage algorithm that includes the  generalized the bilinear message passing for matrix factorization and the   Riemannian manifold gradient-based algorithm for matrix completion. We provided experimental evidences that the proposed approach achieves an accurate cascaded channel estimation for the LIM-assisted massive MIMO systems.

\renewcommand\refname{References}
\bibliographystyle{IEEEtran}
\bibliography{bib-he}

\begin{thebibliography}{10}
\providecommand{\url}[1]{#1}
\csname url@samestyle\endcsname
\providecommand{\newblock}{\relax}
\providecommand{\bibinfo}[2]{#2}
\providecommand{\BIBentrySTDinterwordspacing}{\spaceskip=0pt\relax}
\providecommand{\BIBentryALTinterwordstretchfactor}{4}
\providecommand{\BIBentryALTinterwordspacing}{\spaceskip=\fontdimen2\font plus
\BIBentryALTinterwordstretchfactor\fontdimen3\font minus
  \fontdimen4\font\relax}
\providecommand{\BIBforeignlanguage}[2]{{%
\expandafter\ifx\csname l@#1\endcsname\relax
\typeout{** WARNING: IEEEtran.bst: No hyphenation pattern has been}%
\typeout{** loaded for the language `#1'. Using the pattern for}%
\typeout{** the default language instead.}%
\else
\language=\csname l@#1\endcsname
\fi
#2}}
\providecommand{\BIBdecl}{\relax}
\BIBdecl

\bibitem{liaskos2018new}
C.~Liaskos, S.~Nie, A.~Tsioliaridou, A.~Pitsillides, S.~Ioannidis, and
  I.~Akyildiz, ``A new wireless communication paradigm through
  software-controlled metasurfaces,'' \emph{IEEE Commun. Mag.}, vol.~56, no.~9,
  pp. 162--169, Sep. 2018.

\bibitem{hu2018beyond1}
S.~Hu, F.~Rusek, and O.~Edfors, ``Beyond massive {MIMO}: The potential of data
  transmission with large intelligent surfaces,'' \emph{IEEE Trans. Signal
  Process.}, vol.~66, no.~10, pp. 2746--2758, May 2018.

\bibitem{di2019smart}
\BIBentryALTinterwordspacing
M.~Di~Renzo, M.~Debbah, D.-T. Phan-Huy \emph{et~al.} (Mar. 2019) Smart radio
  environments empowered by {AI} reconfigurable meta-surfaces: {An} idea whose
  time has come. [Online]. Available: \url{https://arxiv.org/abs/1903.08925}
\BIBentrySTDinterwordspacing

\bibitem{wu2018intelligent}
\BIBentryALTinterwordspacing
Q.~Wu and R.~Zhang. (Oct. 2018) Intelligent reflecting surface enhanced
  wireless network via joint active and passive beamforming. [Online].
  Available: \url{https://arxiv.org/abs/1810.03961}
\BIBentrySTDinterwordspacing

\bibitem{huang2018large}
C.~{Huang}, A.~{Zappone}, G.~C. {Alexandropoulos}, M.~{Debbah}, and C.~{Yuen},
  ``Reconfigurable intelligent surfaces for energy efficiency in wireless
  communication,'' \emph{IEEE Trans. Wireless Commun.}, vol.~18, no.~8, pp.
  4157--4170, Aug. 2019.

\bibitem{sleasman2017experimental}
T.~Sleasman, M.~Boyarsky, L.~Pulido-Mancera, T.~Fromenteze, M.~F. Imani, M.~S.
  Reynolds, and D.~R. Smith, ``Experimental synthetic aperture radar with
  dynamic metasurfaces,'' \emph{IEEE Trans. Antennas Propag.}, vol.~65, no.~12,
  pp. 6864--6877, Dec. 2017.

\bibitem{wu2018range}
Z.~Wu, L.~Zhang, H.~Liu, and N.~Kou, ``Range decoupling algorithm for
  accelerating metamaterial apertures-based computational imaging,'' \emph{IEEE
  Sensors J.}, vol.~18, no.~9, pp. 3619--3631, May 2018.

\bibitem{yan2019}
\BIBentryALTinterwordspacing
W.~Yan, X.~Kuai, and X.~Yuan. (May 2019) Passive beamforming and information
  transfer via large intelligent metasurface. [Online]. Available:
  \url{https://arxiv.org/abs/1905.01491}
\BIBentrySTDinterwordspacing

\bibitem{parker2014bilinear}
J.~T. Parker, P.~Schniter, and V.~Cevher, ``Bilinear generalized approximate
  message passing--{Part I}: Derivation,'' \emph{IEEE Trans. Signal Process.},
  vol.~62, no.~22, pp. 5839--5853, Nov. 2014.

\bibitem{wei2016guarantees}
K.~Wei, J.-F. Cai, T.~F. Chan, and S.~Leung, ``Guarantees of {Riemannian}
  optimization for low rank matrix recovery,'' \emph{SIAM J. Matrix Anal.
  Appl.}, vol.~37, no.~3, pp. 1198--1222, Sep. 2016.

\bibitem{taha2019enabling}
\BIBentryALTinterwordspacing
A.~Taha, M.~Alrabeiah, and A.~Alkhateeb. (Apr. 2019) Enabling large intelligent
  surfaces with compressive sensing and deep learning. [Online]. Available:
  \url{https://arxiv.org/abs/1904.10136}
\BIBentrySTDinterwordspacing

\bibitem{zhang2018blind}
J.~Zhang, X.~Yuan, and Y.-J.~A. Zhang, ``Blind signal detection in massive
  {MIMO}: Exploiting the channel sparsity,'' \emph{IEEE Trans. Commun.},
  vol.~66, no.~2, pp. 700--712, Feb. 2018.

\bibitem{liu2018super}
H.~Liu, X.~Yuan, and Y.-J.~A. Zhang, ``Super-resolution blind
  channel-and-signal estimation for massive {MIMO} with one-dimensional antenna
  array,'' \emph{IEEE Trans. Signal Process.}, vol.~67, no.~17, pp. 4433--4448,
  Sep. 2019.

\bibitem{gao2015spatially}
Z.~Gao, L.~Dai, Z.~Wang, and S.~Chen, ``Spatially common sparsity based
  adaptive channel estimation and feedback for {FDD} massive {MIMO},''
  \emph{IEEE Trans. Signal Process.}, vol.~63, no.~23, pp. 6169--6183, Dec.
  2015.

\bibitem{huang2018iterative}
C.~Huang, L.~Liu, C.~Yuen, and S.~Sun, ``Iterative channel estimation using
  {LSE} and sparse message passing for {mmWave} {MIMO} systems,'' \emph{IEEE
  Trans. Signal Process.}, vol.~67, no.~1, pp. 245--259, Jan. 2019.

\bibitem{pi2011introduction}
Z.~Pi and F.~Khan, ``An introduction to millimeter-wave mobile broadband
  systems,'' \emph{IEEE Commun. Mag}, vol.~49, no.~6, pp. 101--107, Jun. 2011.

\bibitem{green1963general}
R.~B. Green, ``The general theory of antenna scattering,'' Ph.D. dissertation,
  The Ohio State University, 1963.

\bibitem{tse2005fundamentals}
D.~Tse and P.~Viswanath, \emph{Fundamentals of wireless communication}.\hskip
  1em plus 0.5em minus 0.4em\relax Cambridge, UK: Cambridge university press,
  2005.

\bibitem{wang2014tens}
P.~Wang, Y.~Li, X.~Yuan, L.~Song, and B.~Vucetic, ``Tens of gigabits wireless
  communications over {E}-band {LoS} {MIMO} channels with uniform linear
  antenna arrays,'' \emph{IEEE Trans. Wireless Commun.}, vol.~13, no.~7, pp.
  3791--3805, Jul. 2014.

\bibitem{aharon2006k}
M.~Aharon, M.~Elad, and A.~M. Bruckstein, ``{K-SVD}: An algorithm for designing
  overcomplete dictionaries for sparse representation,'' \emph{IEEE Trans.
  Signal Process.}, vol.~54, no.~11, pp. 4311--4323, Nov. 2006.

\bibitem{mairal2010online}
J.~Mairal, F.~Bach, J.~Ponce, and G.~Sapiro, ``Online learning for matrix
  factorization and sparse coding,'' \emph{J. Mach. Learn. Res.}, vol.~11, pp.
  19--60, Mar. 2010.

\bibitem{cai2010singular}
J.-F. Cai, E.~J. Cand{\`e}s, and Z.~Shen, ``A singular value thresholding
  algorithm for matrix completion,'' \emph{SIAM J. Optimiz.}, vol.~20, no.~4,
  pp. 1956--1982, Mar. 2010.

\end{thebibliography}

\end{document}